# Upward Point Set Embeddability for Convex Point Sets is in $P^\star$

**(Accepted to GD 2011)**

Michael Kaufmann[1], Tamara Mchedlidze[2], Antonios Symvonis[2]

[1] Wilhelm-Schickard-Institut für Informatik, Universität Tübingen, Germany.
mk@informatik.uni-tuebingen.de
[2] Dept. of Mathematics, National Technical University of Athens, Greece.
{mchet,symvonis}@math.ntua.gr

**Abstract.** In this paper, we present a polynomial dynamic programming algorithm that tests whether a $n$-vertex directed tree $T$ has an upward planar embedding into a convex point-set $S$ of size $n$. Further, we extend our approach to the class of outerplanar digraphs. This nontrivial and surprising result implies that any given digraph can be efficiently tested for an upward planar embedding into a given convex point set.

## 1 Introduction

A *planar straight-line embedding* of a graph $G$ into a point set $S$ is a mapping of each vertex of $G$ to a distinct point of $S$ and of each edge of $G$ to the straight-line segment between the corresponding end points so that no two edges cross each other. Planar straight-line embeddings for outerplanar graphs and trees were studied by Gritzmann *et al.* [11], Bose [4] and Bose *et al.* [5]. Cabello [6] proved that the problem to decide whether a given planar graph admits a planar straight-line embedding into a given point set is $\mathcal{NP}$-hard. Planar graph embeddings into point sets, where edges are allowed to bend, have also been studied (see, e.g., [2,7,12,14,16]).

An *upward planar directed graph* is a digraph that admits a planar drawing such that each edge is represented by a curve monotonically increasing in the $y$-direction. An *upward straight-line embedding* (*UPSE* for short) of an upward planar digraph $G$ into a point set $S$ is a mapping of each vertex of $G$ to a distinct point of $S$ and of each edge to the straight-line segment between its corresponding end points such that no two edges cross and for each edge $(u,v)$ the condition $y(u) < y(v)$ holds. *Upward point set embeddability* is the decision problem of whether a given digraph has an UPSE into a given point set.

$^\star$ This research has been co-financed by the European Union (European Social Fund - ESF) and Greek national funds through the Operational Program "Education and Lifelong Learning" of the National Strategic Reference Framework (NSRF) - Research Funding Program: Heracleitus II. Investing in knowledge society through the European Social Fund.

Upward point set embeddability was first studied by Giordano et al. [9]. The authors studied the version of the problem where bends on edges are allowed and showed that every planar $st$-digraph admits an upward point set embedding with at most two bends per edge. Upward point set embeddability with a given mapping, i.e., where a correspondence between the nodes and the point set is part of the input, was studied in [10,15]. Recently, straight-line drawings were studied in [1,3,8] and many interesting and partial results were presented. Among them are several results concerning upward point set embeddability of a tree into a convex point set. More specifically, several families of trees were presented, which have an UPSE into every convex point set, i.e., caterpillars, switch-trees, hourglass trees. On the other hand, it was demonstrated that the family of $k$-switch trees (generalization of switch-trees) does not have an UPSE into all convex point sets. An immediate question that arises from these facts is whether the existence of an UPSE of a tree into a convex point set can be efficiently tested. The contribution of this paper is an affirmative answer to this question. More specifically, we show that, given a directed tree $T$ and a convex point set $S$, it can be tested in polynomial time whether $T$ has an UPSE into $S$.

Recently, Geyer *et al.* [8] proved that the general *upward point-set embeddability* problem is $\mathcal{NP}$-complete even for $m$-convex point sets[3]. Thus one interesting open problem regarding UPSE was whether there exists a class of upward planar digraphs $\mathcal{D}$ for which the decision problem whether a digraph $D \in \mathcal{D}$ admits an UPSE into a given point set $S$ remains $\mathcal{NP}$-complete even for a convex point set $S$. We answer this question in the negative by extending our UPSE algorithm for trees to the class of outerplanar graphs. Since any graph admitting a planar embedding into a convex point set is an outerplanar digraph, our result implies that the upward point-set embeddability can be efficiently solved for convex point sets and general digraphs.

For simplicity of presentation, we first concentrate on the case of directed trees. In Section 2, we present the necessary notation and some basic results on UPSE, which are utilized by our tree algorithm. In Section 3, we study a restricted version of the UPSE problem which fixes the point in which the root of the tree is embedded and places restrictions on the drawing of subtrees. In Section 4, we present a dynamic programming algorithm for deciding whether a directed tree has an UPSE into a convex point set. In Section 5 we state the extended result for outerplanar digraphs. Due to space constraints we present all the material concerning outerplanar digraphs in the Appendix.

## 2 Notation - Preliminaries

**Point sets.** Let $S$ be a set of points on the plane. We assume that the points of $S$ are in general position, i.e., no three of them lie on the same line. Moreover, we also assume that no two points of $S$ share the same $y$-coordinate; if they do, a slight rotation of the coordinate axes can ensure that all points have distinct $y$-coordinates. The *convex hull* $CH(S)$ of $S$ is the point set that is obtained as a convex combination of the points of $S$. A point set such that no point is in the convex hull of the others is called a *point set in convex position*, or a *convex point set*. Given a point set $S$, by $t(S)$ (resp., $b(S)$) we denote the top (bottom) point of $S$ i.e., the point with the largest (resp., smallest) $y$-coordinate.

---

[3] An $m$-convex point set can be intuitively defined as a set of $m$ shelled, one into another, distinct convex point sets.

A *one-sided convex point set* $S$ is a convex point set in which $b(S)$ and $t(S)$ are adjacent on the border of $CH(S)$. If $t(S)$ and $b(S)$ appear adjacent and in this order on the border of $CH(S)$ as we traverse it in the clockwise (resp., counterclockwise) direction, then the one-sided convex point set is called a *left-sided convex point set* (resp., *right-sided convex point set*). A point set consisting of at most two points is considered to be either a left-sided or a right-sided convex point set. A convex point set which is not one-sided, is called a *two-sided convex point set*.

Each given convex point set $S$ may be considered to be the union of two specified (at the time $S$ is given) one-sided convex point sets, one left-sided which is denoted by $L(S)$ and is referred to as the *left-side* of $S$, and one right-sided which is denoted by $R(S)$ and is referred to as the *right-side* of $S$. When there is no confusion regarding the point set $S$ we refer to, for simplicity, we use the terms $L$ and $R$ instead of $L(S)$ and $R(S)$, respectively. Each of the points $b(S)$ and $t(S)$ belongs to either $L(S)$ or $R(S)$ but not both.

A subset of points of a convex point set $S$ is called *consecutive* if its points appear consecutively as we traverse the convex hull of $S$ in clockwise direction. Given that all points of $S$ have distinct $y$-coordinates, we can refer to the first, the second, the third, etc., lowest point on the left (right) side of $S$. By $p_i^L$, $1 \leq i \leq |L(S)|$, we denote the $i$-th lowest point on the left side of $S$. Similarly, by $p_i^R$, $1 \leq i \leq |R(S)|$, we denote the $i$-th lowest point on the right side of $S$.

Let $S_{a..b,c..d} = \{p_i^L \mid a \leq i \leq b\} \cup \{p_i^R \mid c \leq i \leq d\}$ denote the subset of $S$ consisting of $b - a + 1$ consecutive points on the left side of $S$, starting from point $p_a^L$ in the clockwise direction, and of $d-c+1$ consecutive points on the right side, starting from point $p_c^R$ in the counterclockwise direction. For simplicity, for a one-sided point set $S$ we use the notation $S_{a..b}$.

In this paper, we assume that queries of the form *"Find the $i$-th point on the left/right side of the convex point set $S$"* can be answered in $O(1)$ time, e.g., the points on each side of $S$ are stored in an array in ascending order of their $y$-coordinates.

**Trees.** Consider a *directed tree* $T$, i.e., a directed acyclic graph whose underlying undirected structure is that of a tree. Tree $T$ is *rooted* if one of its vertices, denoted by $r(T)$, is designated as its *root*. We then say that $T$ *is rooted at* vertex $r(T)$. By $d^-(v)$ (resp., $d^+(v)$) we denote the in-degree (resp., the out-degree) of vertex $v$ of $T$. By $d(v)$ we denote the total degree of vertex $v$, i.e., $d(v) = d^-(v) + d^+(v)$.

Let $T$ be a rooted tree and let $r = r(T)$ be its root. Let $T_1^l, \ldots, T_{d^-(r)}^l, T_1^h, \ldots, T_{d^+(r)}^h$ be the rooted subtrees of $T$ obtained by removing from $T$ its root $r$ and $r$'s inci-dent arcs and having as their roots the vertices that are incident to $r$ by either an incoming or an outgoing arc (see Figure 1.a). Trees $T^l_1, \ldots, T^l_{d^-(r)}, T^h_1, \ldots, T^h_{d^+(r)}$ are called the *subtrees of* $T$. Note that the superscripts "$l$" and "$h$" indicate whether a particular subtree of $T$ is connected to $r$ by an incoming to $r$ or by an outgoing from $r$ arc, respectively.

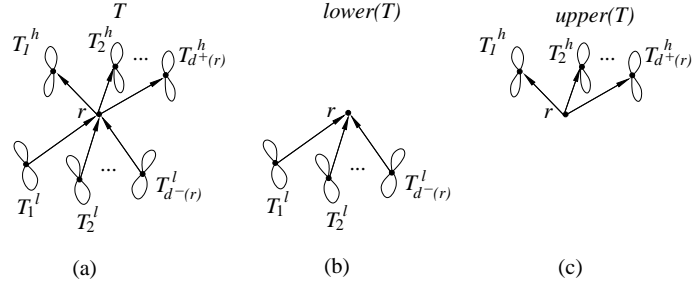

**Fig. 1.** (a) A rooted at vertex $r$ tree $T$ and its subtrees $T^l_1, \ldots, T^l_{d^-(r)}$, $T^h_1, \ldots, T^h_{d^+(r)}$. (b) The subtree $lower(T)$ of $T$. (c) The subtree $upper(T)$ of $T$.

The rooted subtree of $T$ consisting of $T$'s root, $r$, together with $T^l_1, \ldots, T^l_{d^-(r)}$ is called the *lower subtree of* $T$ and is also rooted at $r$. The lower subtree of $T$ is denoted by $lower(T)$ (Figure 1.b). Similarly, the rooted subtree of $T$ consisting of $T$'s root, $r$, together with $T^h_1, \ldots, T^h_{d^+(r)}$ is called the *upper subtree of* $T$ and is also rooted at $r$. The upper subtree of $T$ is denoted by $upper(T)$ (Figure 1.c). In this paper, we use the notation $\{u, v\}$ to denote arc $(u, v)$ if $(u, v) \in T$ or arc $(v, u)$ if $(v, u) \in T$. If $u$ is mapped to point $p$ and $v$ is mapped to point $q$ that is located below $p$, then we say that $\{u, v\}$ is drawn upwards (downwards) if $(v, u) \in T$ ($(u, v) \in T$).

### 2.1 Some known results on UPSE of rooted directed trees

We present some known results on UPSE of rooted directed trees that will be utilized by our algorithms. Binucci *et al.*[3] proved the following lemma concerning the placement of the subtrees of $T$ in an UPSE of $T$ on a convex point set.

**Lemma 1 (Binucci *et al.* [3]).** *Let $T$ be a n-vertex directed tree rooted at $r$ and let $S$ be any convex point set of size $n$. Let $T_1, T_2, \ldots, T_{d(r)}$ be the subtrees of $T$. Then, in any UPSE of $T$ into $S$, the vertices of subtree $T_i$ are mapped to a set of consecutive points of $S$, $1 \leq i \leq d(r)$.* □

The following lemma concerns the UPSE of a rooted tree into a *one-sided* convex point set. It can be considered to be a simple restatement of a result by Heath *et al.* [13] (Theorem 2.1).

**Lemma 2.** *Let $T$ be a $n$-vertex directed tree rooted at $r$ and $S$ be a one-sided convex point set of size $n$. Let $T_1, T_2, \ldots, T_{d(r)}$ be the subtrees of $T$. Then, $T$ admits an UPSE into $S$ so that the following are true:*

  i) *Each $T_i$, $1 \leq i \leq d(r)$, is drawn on consecutive points of $S$.*
  ii) *If the root $r$ of $T$ is mapped to point $p_r$ then there is no arc connecting a point of $S$ below $p_r$ to a point of $S$ above $p_r$.*

By utilizing Lemma 2, we prove the following.

**Lemma 3.** *Let $T$ be a $n$-vertex directed tree rooted at $r$ and $S$ be a one-sided convex point set of size $n$. Then, an UPSE of $T$ into $S$ satisfying the properties of Lemma 2 can be obtained in $O(n)$ time. Moreover, after $O(n)$ time preprocessing, the point $p_r$ that hosts the root $r$ of $T$ can be determined in $O(1)$ time (i.e., without determining the complete UPSE of $T$ into $S$).*

*Proof.* Let $k = |lower(T)|$ be the size of subtree $lower(T)$ (rooted at $r$). It immediately follows that in an UPSE of $T$ into $S$ satisfying the properties of Lemma 2 there are $k - 1$ vertices of $T$ (all belonging to $lower(T)$) that are placed below $r$. Thus, $r$ is mapped to the $k$-th lowest point of $S$. This point, say $p_r$, can be computed in $O(1)$ time. Having decided where to place the root $r$, the UPSE of $T$ can be completed in $O(n)$ time by recursively embedding the vertices of $lower(T)$ ($upper(T)$) to the points of $S$ below (above) $p_r$. □

## 3 A restricted UPSE problem for rooted directed trees

In this section, we study a restricted UPSE problem that will be later on used by our main algorithm which decides whether there exists an UPSE of a given directed tree into a given convex point set.

**Definition 1.** *In a restricted UPSE problem for trees we are given a directed tree $T$ rooted at $r$, a convex point set $S$, and a point $p_r \in S$. We are asked to decide whether there exists an UPSE of $T$ into $S$ such that (i) the root $r$ of $T$ is mapped to point $p_r$ and, (ii) each subtree of $T$ (rooted at $r$) is mapped to consecutive points on the same side (either $L$ or $R$) of $S$.*

The following observation follows directly from the definition of a restricted UPSE.

**Observation 1** *In a restricted UPSE of a directed tree $T$ rooted at $r$ into a convex point set $S$, where the root $r$ of $T$ is mapped to point $p_r \in S$, no edge enters triangles $\triangle(t(L), t(R), p_r)$ and $\triangle(b(L), b(R), p_r)$.*

Figure 2.a shows a tree $T$ rooted at vertex $r$, a convex point set $S$ consisting of a left-sided convex point set $L$ and a right-sided convex point set $R$. Tree $T$ has a restricted UPSE only if its root $r$ is mapped to point $p_r \in L$ (Figure 2.b).

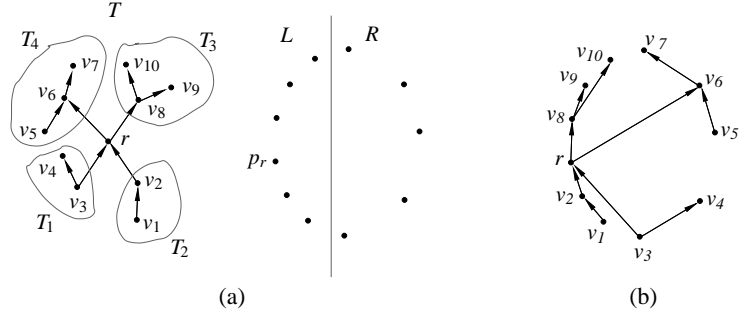

**Fig. 2.** (a) A tree $T$ rooted at vertex $r$ and a convex point set $S = L \cup R$. (b) A restricted UPSE of $T$ into $S$ so that $r$ is mapped to point $p_r$. No restricted UPSE of $T$ exists when $r$ is mapped to any point other than $p_r$.

Mapping $r$ to any other point $p \in S$ makes it impossible to map each subtree of $T$ to consecutive points on the same side of $S$.

Before we proceed to describe a decision algorithm for the restricted UPSE problem, we need some more notation. Let $T$ be a directed tree rooted at vertex $r$ and let $\lambda = (T_1, \ldots, T_{d(r)})$ be an ordering of the subtrees of $T$. Let $S$ be a convex point set and let $\Gamma$ be an UPSE of $T$ into $S$. We say that UPSE $\Gamma$ *respects ordering* $\lambda$ if for any two subtrees $T_i$ and $T_j$, $1 \leq i \leq j \leq d(r)$, that are both mapped on the same side of $S$, $T_i$ is mapped to a point set that is entirely below the point set $T_j$ is mapped to.

Consider a tree $T$ rooted at vertex $r$ and let $\lambda = (T_1^l, \ldots, T_{d^-(r)}^l, T_1^h, \ldots, T_{d^+(r)}^h)$ be an ordering of the subtrees of $T$. Ordering $\lambda$ is called a *proper ordering* of the subtrees of $T$ if it satisfies the following properties:

(i) $|upper(T_i^l)| \leq |upper(T_j^l)|$, $1 \leq i \leq j \leq d^-(r)$, and
(ii) $|lower(T_i^h)| \geq |lower(T_j^h)|$, $1 \leq i \leq j \leq d^+(r)$.

For example, ordering $\lambda_1 = (T_2, T_1, T_4, T_3)$ is a proper ordering of the subtrees of $T$ in Figure 2.a since $|upper(T_2)| < |upper(T_1)|$ and $|lower(T_4)| > |lower(T_3)|$ while ordering $\lambda_2 = (T_1, T_2, T_3, T_4)$ is not. Observe that in a proper ordering $\lambda$ of $T$, the subtrees in the lower subtree of $T$ appear before the subtrees in the upper subtree of $T$. The proof of the following lemma can be found in Appendix A.

**Lemma 4.** *Let $T$ be a n-vertex directed tree rooted at vertex $r$, $\lambda$ be a proper ordering of the subtrees of $T$, and $S$ be a convex point set of size $n$. Then, if there exists a restricted UPSE of $T$ into $S$, there also exists a restricted UPSE of $T$ into $S$ that respects $\lambda$.* □

**Theorem 1.** *Let $T$ be a n-vertex directed tree rooted at vertex $r$, $L$ and $R$ be left-sided and right-sided convex point sets, resp., such that $S = L \cup R$ is a convex point set of size $n$, and $p_r$ a point of $S$. The restricted UPSE problem with input* $T$, $S$ and $p_r$ can be decided in $O(d(r)n)$ time. Moreover, if a restricted UPSE for $T$, $S$ and $p_r$ exists, it can also be constructed in $O(d(r)n)$ time.

*Proof.* Let $\lambda = (T_1, T_2, \ldots, T_{d(r)})$ be a proper ordering of the subtrees of $T$. Proper ordering $\lambda$ can be computed in $O(n)$ time by a simple tree traversal that computes at the root of $T$ the number of vertices in each subtree of $T \setminus \{v\}$ followed by a bucket sort of the sizes of the subtrees rooted at $r$. Since the restricted UPSE problem will be repeatedly solved on subtrees of $T$, we assume that $T$ has been appropriately preprocessed in $O(n)$ time and, thus, a proper ordering of these subtrees can be then computed in $O(d(r))$ time. By Lemma 4, it is enough to test whether there exists a restricted UPSE that respects $\lambda$. Thus, we will describe a dynamic programming algorithm that tests whether there exists a restricted UPSE on input $T$, $L$, $R$ and $p_r$.

Our dynamic programming algorithm uses a two-dimensional $d(r) \times |L|$ matrix $M$. Value $M[i,j]$ is $TRUE$ if and only if there exists a restricted UPSE of the subtree of $T$ induced by $r$ and $T_1, \ldots, T_i$ that uses all the $j$ lowest points of the left-sided point set $L$ and as many consecutive points as required in the lowest part of the right-sided convex point set $R$. Recall that $\{u,v\}$ denotes arc $(u,v)$ if $(u,v) \in T$; arc $(v,u)$ if $(v,u) \in T$; otherwise it is undefined.

For the boundary conditions of our dynamic programming we have that:
$M[0,0] = TRUE$
$$M[1,j] = \begin{cases} TRUE, & \text{if } j=0 \text{ and } p_r \notin R_{1..|T_1|} \text{ and } \{r(T_1), p_r\} \text{ is upward} \\ TRUE, & \text{if } j=|T_1| \text{ and } p_r \notin L_{1..|T_1|} \text{ and } \{r(T_1), p_r\} \text{ is upward} \\ FALSE, & \text{otherwise} \end{cases}$$
Let $\sigma = |T_1| + \ldots + |T_i|$. $M[i,j]$, $1 < i \leq d(r)$ and $0 \leq j \leq |L|$, is set to $TRUE$ if any of the following conditions is true; otherwise it is set to $FALSE$.

**c-1:** $M[i, j-1] = TRUE$ **and** $p_r = L_{j..j}$.
This is the case where point $p_r$ happens to be the $j$-th point of $L$. There is no need to test for upwardness of $\{r(T_i), p_r\}$ since it has been already tested when entry $M[i, j-1]$ was filled in.

**c-2:** $M[i-1, j-|T_i|] = TRUE$ **and** $p_r \notin L_{j-|T_i|+1..j}$ **and** $\{r(T_i), p_r\}$ *is upward*.
In this case, $T_i$ is placed on $L$. We know that $T_i$ fits on $L$ since $j < |L|$, however, we must make sure that it also holds that $p_r$ is not one of the $|T_i|$ topmost points of $L_{1..j}$.

**c-3:** $M[i-1, j] = TRUE$ **and** $p_r \in R_{1..\sigma-j-|T_i|+1}$ **and** $\sigma - j + 1 \leq |R|$ **and** $\{r(T_i), p_r\}$ *is upward*.
In this case, $T_i$ is placed to $R$. If $p_r$ is one of the points in $L_{1..\sigma-j-|T_i|+1}$ then we have to make sure that at least $\sigma - j + 1$ points exist in $|R|$.

**c-4:** $M[i-1, j] = TRUE$ **and** $p_r \notin R_{1..\sigma-j}$ **and** $\sigma - j \leq |R|$ **and** $\{r(T_i), p_r\}$ *is upward*.
In this case, $T_i$ is also placed to $R$. However, in contrast to case **c-3**, $p_r$ is not one of the points in $L_{1..\sigma-j}$. Thus, we only need to make sure that at least $\sigma - j$ points exist in $|R|$.

When determining the value of an entry $M[i,j]$ we need to decide whether arc $\{r(T_i), p_r\}$ is upward. In order to do that, we need to know the point to which $r(T_i)$ is mapped. By Lemma 3, this point can be computed in $O(1)$ time since $T_i$ is mapped to $|T_i|$ consecutive points forming a one-sided convex point set.

It can be easily verified that entry $M[d(r), |L|] = TRUE$ if and only if there is a restricted UPSE of $T$ into $L \cup R$ such that $r(T)$ is mapped to $p_r$.

Each entry of matrix $M$ can be filled in $O(1)$ time. Thus, all entries of matrix $M$ are filled in $O(d(r)|L|)$ time. In the event that a restricted UPSE of $T$ into $L \cup R$ such that $r(T)$ is mapped to $p_r$ exists, we can construct the embedding by storing in each entry $M[i,j]$ that was set to $TRUE$ the side ("L" or "R") in which $T_i$ was placed. This information, together with the fact that the restricted UPSE respects ordering $\lambda$ is sufficient to construct the embedding. □

Denote by $\mathcal{L}(T, L, R)$ the set of points $p \in L \cup R$ such that there exists a restricted UPSE of $T$ on $L \cup R$ where the root of $T$ is mapped to $p$. The next theorem follows easily from Theorem 1 by testing each point of $L \cup R$ as a candidate host for $r(T)$.

**Theorem 2.** *Let $T$ be an n-vertex directed tree rooted at vertex $r$ and $L$ and $R$ be left-sided and right-sided convex point sets, resp., such that $S = L \cup R$ is a convex point set of size $n$. Then, the set $\mathcal{L}(T, L, R)$ can be computed in $O(d(r)n^2)$ time.* □

**Note:** In this paper we only consider embeddings of $n$-vertex trees into point sets of size $n$. Thus, by definition $\mathcal{L}(T, L, R)$ is empty when $|T| \neq |L| + |R|$.

## 4  The testing algorithm for directed trees

Let $T$ be a directed tree and let $S$ be a convex point set. In any UPSE of $T$ into $S$, a source node $s$ and a sink node $t$ of $T$ will be mapped to points $b(S)$ and $t(S)$, respectively. In this section, we present a dynamic programming algorithm that decides in polynomial time whether, given a $n$-vertex directed tree $T$, a source $s$ and a sink $t$ of $T$, and a convex point set $S$ of size $n$, $T$ has an UPSE on $S$ so that $s$ and $t$ are mapped to $b(S)$ and $t(S)$, respectively. Applying this algorithm on all $\langle source, sink \rangle$ pairs of $T$, yields a polynomial time algorithm for deciding whether $T$ has an UPSE on $S$.

Let $s$ and $t$ be a source and a sink vertex of $T$, respectively. Denote by $P_{s,t} = \{s = w_1, w_2, \ldots, w_m = t\}$ the (undirected) path connecting $s$ and $t$ in $T$, see Figure 3.a. By $T_{s,w_i}$, $1 \leq i < m$, we denote the subtree of $T$ that contains source $s$ and is formed by the removal of edge $\{w_i, w_{i+1}\}$. By definition, we set $T_{s,w_m} = T$. Let $T_{w_i} = T_{s,w_i} \setminus T_{s,w_{i-1}}$, $1 < i \leq m$. By definition, $T_{w_1} = T_{s,w_1}$. By Lemma 1, we know that $T_{s,w_i}$ is drawn on consecutive points of $S$, call this point set $S_i$ (see also Figure 3.b). Since $s$ is mapped to $b(S)$, we infer that $b(S) \in S_i$. Similarly, in any UPSE of $T$ into $S$, $T_{s,w_{i+1}}$ is also drawn on consecutive points of $S$ that contain $b(S)$, call this point set $S_{i+1}$. Hence, $T_{w_{i+1}}$ is drawn on a set $S_{w_{i+1}} = S_{i+1} \setminus S_i$, that is, a subset of $S$ comprised by two consecutive point sets of $S$, one on its left and one on its right side.

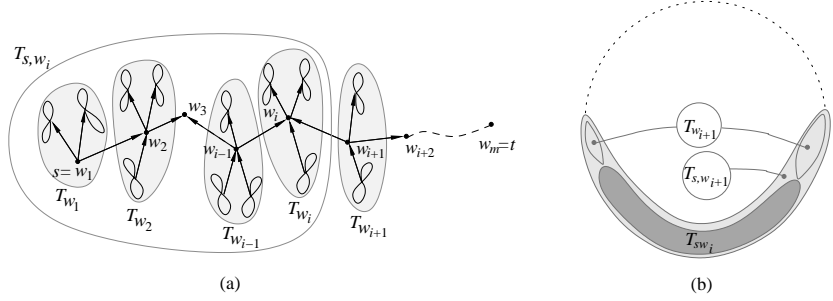

**Fig. 3.** (a) The decomposition of tree $T$ based on a path between a source $s$ and a sink $t$ of $T$. (b) The structure of an UPSE of the tree $T$ into point set $S$.

Our dynamic programming algorithm maintains a list of points $\mathcal{P}(a,b,k)$, $0 \leq a \leq |L|$, $0 \leq b \leq |R|$, $1 \leq k \leq m$, such that:

$$p \in \mathcal{P}(a,b,k) \iff \begin{cases} T_{s,w_k} \text{ has an UPSE into point set } S_{1..a,1..b} \text{ with} \\ \text{vertex } w_k \text{ mapped to point } p. \end{cases}$$

For the boundary conditions of our dynamic programming we have that:

$\mathcal{P}(a,b,1) = \mathcal{L}(T_{w_1}, L_{1..a}, R_{1..b})$ where $a + b = |T_{w_1}|$.

Note that since $w_1$ is a source, $\mathcal{P}(a,b,1)$ is either $\{b(s)\}$ or $\emptyset$.

Our dynamic programming is based on the following recurrence relation, which allows us to add points in $\mathcal{P}(a,b,i)$. For any $1 < i \leq m$ we set:

$\mathcal{P}(a,b,i) = \{p \mid \exists a_1, b_1 \in Z : a_1 + b_1 = |T_{w_i}|$
$\quad\quad\quad$ **and** $p \in \mathcal{L}(T_{w_i}, L_{a-a_1+1..a}, R_{b-b_1+1..b})$
$\quad\quad\quad$ **and** $\exists q \in \mathcal{P}(a-a_1, b-b_1, i-1)$
$\quad\quad\quad$ **and** $\{p,q\}$ is upward $\}$

We first prove that if $p \in \mathcal{P}(a,b,i)$ then $T_{s,w_i}$ has an UPSE into point set $S_{1..a,1..b}$ with vertex $w_i$ mapped to point $p$. Assume that $\mathcal{P}(a-a_1, b-b_1, i-1) \neq \emptyset$ and let $q$ be a point in it. Thus, $T_{s,w_{i-1}}$ has an UPSE with vertex $w_{i-1}$ mapped to point $q$. Also assume that $\mathcal{L}(T_{w_i}, L_{a-a_1+1..a}, R_{b-b_1+1..b}) \neq \emptyset$ and let $p$ be a point in it. Thus, $T_{w_i}$ has a restricted UPSE in $L_{a-a_1+1..a} \cup R_{b-b_1+1..b}$ with $w_i$ mapped to $p$. If arc $\{w_{i-1}, w_i\}$ which is drawn as line-segment $(q,p)$ is upward, then we can combine the UPSE for $T_{s,w_{i-1}}$ with the restricted UPSE for $T_{w_i}$ in order to get an UPSE of $T_{s,w_i}$ on point set $S_{1..a,1..b}$. Note that, by Observation 1, we have that the combined drawing is planar. Thus, we conclude that point $p$ belongs to $\mathcal{P}(a,b,i)$.

For the reversed statement we also work by induction. From the boundary conditions we know that if $T_{s,w_1} = T_{w_1}$ has an UPSE in to a point set $S_{1..a,1..b}$ then $b(S) \in \mathcal{P}(a,b,1)$, where $a + b = |T_{w_1}|$. Assume that the statement is true

**Algorithm 1:** TREE-UPSE($T, S, s, t$)

**input** : A directed tree $T$, a point set $S$, a source $s$ and a sink $t$ of $T$. Path $(s = w_1, \ldots, w_m = t)$ is used to progressively build tree $T$ from subtrees $T_{w_i}$, $1 \leq i \leq m$.

**output** : "YES" if $T$ has an UPSE into $S$ with $s$ mapped to $b(S)$ and $t$ mapped to $t(S)$, "NO" otherwise.

1. **For** $a = 0 \ldots |L|$
2.    **For** $b = 0 \ldots |R|$
3.       $\mathcal{P}(a,b,1) = \mathcal{L}(T_{w_1}, L_{1..a}, R_{1..b})$
4.       **For** $k = 2 \ldots m$       //Consider tree $T_{w_k}$
5.          $\mathcal{P}(a,b,k) = \emptyset$
6.          **For** $i = 0 \ldots |T_{w_k}|$    //We consider the case where $i$ vertices of $T_{w_k}$ are placed to the left side of $S$
7.             **if** $(a - i \geq 0)$ **and** $(b - (|T_{w_k}| - i) \geq 0)$
8.                Let $\mathcal{L} = \mathcal{L}(T_{w_k}, L_{a-i+1..a}, R_{b-(|T_{w_k}|-i)+1..b})$
9.                //We consider all possible placements of $w_{k-1}$
10.                **For each** $q$ in $\mathcal{P}(a-i, b-(|T_{w_k}|-i), k-1)$
11.                    //We consider all the possible placements of vertex $w_k$
12.                    **For each** $p$ in $\mathcal{L}$
13.                        **if** ( $\{w_{i-1}, w_i\}$ drawn on line-segment $(q, p)$ is upward )
14.                           **then** add $p$ to $\mathcal{P}(a,b,k)$.
15. **if** $\mathcal{P}(|L|, |R|, m)$ is empty **then return**("NO");
16. **return**("YES");

for $T_{s,w_{i-1}}$, i.e., if $T_{w_{i-1}}$ has an UPSE in to a point set $S_{1..a,1..b}$ with vertex $w_{i-1}$ mapped to $q$ then $q \in \mathcal{P}(a, b, i-1)$. Assume also that $T_{w_i}$ has an UPSE in to a point set $S_{1..a,1..b}$ with vertices $s$ and $w_i$ mapped to points $b(S)$ and $p$, respectively. By the discussion above we know that in every such embedding $T_{s,w_{i-1}}$ is mapped to consecutive points of $S_{1..a,1..b}$ that contains $b(S)$. Therefore there exist two numbers $a_1$ and $b_1$, so that $a_1 + b_1 = |T_{w_i}|$ and subtree $T_{w_i}$ is mapped to the point set $S_{a-a_1+1..a,b-b_1+1..b}$, with vertex $w_i$ mapped so some point $p$, $p \in \mathcal{L}(T_{w_i}, L_{a-a_1+1..a}, R_{b-b_1+1..b})$. Moreover, by induction hypothesis, there exists $q \in \mathcal{P}(a-a_1, b-b_1, i-1)$. So, since the edge connecting $p$ and $q$ is upward, by the definition of recurrence relation we infer that $p \in \mathcal{P}(a, b, i)$.

Finally we note that, an UPSE of $T$ into $S$ such that source $s$ and sink $t$ are mapped to $b(S)$ and $t(S)$, respectively, exists if and only if $\mathcal{P}(|L|, |R|, m)$ is non-empty. Note that if $\mathcal{P}(|L|, |R|, m) \neq \emptyset$, then it must hold that $\mathcal{P}(|L|, |R|, m) = \{t(S)\}$.

$\mathcal{P}(a, b, k)$, when $0 \leq a \leq |L|$, $0 \leq b \leq |R|$, $1 \leq k \leq m$ is calculated by Algorithm 1.

**Theorem 3.** *Let $T$ be a n-vertex rooted directed tree, $S$ be a convex point set of size $n$, $s$ be a source of $T$ and $t$ be a sink of $T$. Then, it can be decided in time $O(n^5)$ whether $T$ has an UPSE on $S$ such that $s$ is mapped to $b(S)$ and $t$* *is mapped to $t(S)$. Moreover, if such an UPSE exists, it can also be constructed within the same time bound.*

*Proof.* A naive analysis of Algorithm 1 yields an $O(n^7)$ time complexity. The analysis assumes that (i) the left and the right side of $S$ have both size $O(n)$, (ii) the path from $s$ to $t$ has length $O(n)$, (iii) each tree $T_{w_i}$ has size $O(n)$ and (iv) each $\mathcal{L}$-list containing the solution of a restricted UPSE problem is computed in $O(n^3)$ time. However, based on the following two observations, the total time complexity can be reduced to $O(n^5)$.

- A factor of $n$ can be saved by realizing that in our dynamic programming we can maintain a list $\mathcal{P}'(a,i)$ which uses only one parameter for the left side of the convex set (in contrast with $\mathcal{P}(a,b,i)$ which uses a parameter for each side of $S$). The number of points on the right side of $S$ is implied since the size of each tree $T_{s,w_i}$ is fixed. For simplicity, we have decided to use notation $\mathcal{P}(a,b,i)$.
- Another factor of $n$ can be saved by observing that the solution of a restricted UPSE is actually $O(deg(w_i)n^2)$. Thus, summing over all $i$ gives $O(n^3)$ in total, and not $O(n^4)$.

The UPSE of $T$ into $S$ can be recovered easily by modifying Algorithm 1 so that it stores for each point $p \in \mathcal{P}(a,b,k)$ the point $q$ where vertex $w_{i-1}$ is mapped to as well as the point set that hosts tree $T_{s,w_{i-1}}$ (i.e., its top point on the left and the right side of $S$). □

By applying Algorithm 1 on all $\langle source, sink \rangle$ pairs of $T$ we can decide whether tree $T$ has an UPSE on a convex point set $S$, as the main next theorem indicates.

**Theorem 4.** *Let $T$ be a $n$-vertex rooted directed tree and $S$ be a convex point set of size $n$. Then, it can be decided in time $O(n^6)$ whether $T$ has an UPSE on $S$. Moreover, if such an UPSE exists, it can also be constructed within the same time bound.*

*Proof.* Note that a naive application of the idea leads to the algorithm with time complexity $O(n^7)$, since there are $O(n^2)$ distinct pairs of sources and sinks. Next we explain how the overall time complexity can be reduced to $O(n^6)$. Let $P_{s,t}$ be a path from $s$ to $t$, passing through $m$ vertices, and let $t'$ be the $j$-th vertex of $P_{s,t}$ that is also a sink of $G$. During the computation of $\mathcal{P}(a,b,m)$ corresponding to path $P_{s,t}$ we also compute $\mathcal{P}(a,b,j)$ and thus we can immediately answer whether there exists an UPSE of $G$ into $S$ so that $s$ and $t'$ is mapped to $b(S)$ and $t(S)$, respectively. Next consider a sink $\tilde{t}$ that does not belong to path $P_{s,t}$. Consider the path $P_{s,\tilde{t}}$. Assume that the last common vertex of $P_{s,t}$ and $P_{s,\tilde{t}}$ is the $j$-th vertex of $P_{s,t}$. In order to compute whether there is an UPSE of $G$ into $S$ so that $s$ and $\tilde{t}$ are mapped to $b(S)$ and $t(S)$, respectively, we can start the computations of Algorithm 1 determined by variable $k$ from the $j+1$-th step (see line 4 of the algorithm). Thus, for a single source $s$ and all possible sinks variable $k$ changes $n$ times. Since the number of different sources is $O(n)$ we conclude that the whole algorithm runs in time $O(n^6)$. □

## 5  Generalization to directed graphs

Let $G$ be a general directed graph with $n$ vertices and $S$ be a convex point set of size $n$. A necessary condition for $G$ to admit a planar embedding into $S$ is to be outerplanar. In Appendix B we show how the algorithm which tests whether a directed tree has an UPSE into a convex point set can be extended to the class of outerplanar digraphs. The construction is along the same lines as for trees, but technically more involved. Therefore, we decided to keep the descriptions separated. Summarizing, we obtain the following theorem.

**Theorem 5.** *Let $G$ be a $n$-vertex digraph and $S$ be a convex point set of size $n$. It can be decided in polynomial time whether $G$ has an UPSE on $S$. Moreover, if such an UPSE exists, it can also be constructed in polynomial time.*  □

**Acknowledgments.** We thank Markus Geyer for the useful discussions during the work on this paper.

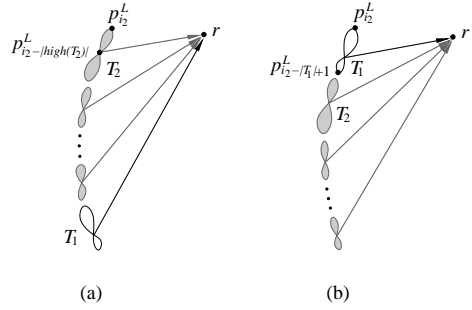

**Fig. 4.** The configuration of subtrees used in the proof of Lemma 4. (a) The drawing of subtrees $T_1$ and $T_2$ does not respect proper ordering $\lambda$. (b) Subtrees $T_1$ and $T_2$, as well as the subtrees placed between them, have been redrawn so that proper ordering $\lambda$ is respected (as far as $T_1$ and $T_2$ are concerned).

# Appendix

## A  Proof of Lemma 4

**Lemma 4.** *Let $T$ be a n-vertex directed tree rooted at vertex $r$, $\lambda$ be a proper ordering of the subtrees of $T$, and $S$ be a convex point set of size $n$. Then, if there exists a restricted UPSE of $T$ into $S$, there also exists a restricted UPSE of $T$ into $S$ that respects $\lambda$.*

*Proof.* Consider a restricted UPSE $\Gamma$ of $T$ into $S$ and assume that it does not respect ordering $\lambda$. Consider any two subtrees $T_1$ and $T_2$ of $T$ that are mapped on the same side of $S$, say both are drawn on the left side of $S$ and $T_1$ is drawn below $T_2$, and assume that they appear in reverse order in $\lambda$.

First observe that both $T_1$ and $T_2$ belong either to the lower or to the upper subtree of $T$. If they do not, and since they do not respect ordering $\lambda$, $T_1$ belongs to the upper subtree of $T$ and $T_2$ in the lower subtree of $T$. Then, it is impossible that edges $(r, r(T_1))$ and $(r(T_2), r)$ that belong to $T$ are both drawn upward in any restricted UPSE of $T$ into $S$ with $T_1$ drawn below $T_2$.

Without lost of generality assume that both $T_1$ and $T_2$ belong to the lower subtree of $T$ (the proof where they both belong to the upper subtree of $T$ is symmetric). Let the highest point of $T_2$ be mapped to the $i_2$-th lowest point on the left side of $S$, i.e., point $p^L_{i_2}$ (see Figure 4.a). Consider the drawing $\Gamma'$ obtained from $\Gamma$ by shifting downwards by $|T_1|$ points the drawing of subtree $T_2$ and of all the subtrees drawn between $T_1$ and $T_2$ in $\Gamma$, and by drawing $T_1$ (as it was drawn in $\Gamma$) at the $|T_1|$ points $\{p^L_{i_2} \ldots p^L_{i_2-|T_1|+1}\}$ (see Figure 4.b). The resulting drawing $\Gamma'$ is obviously planar. In order to prove that $\Gamma'$ is a restricted UPSE it is sufficient to prove that both edges $(r(T_1), r)$ and $(r(T_2), r)$ remain upward. Edge $(r(T_2), r)$ obviously remains upward since vertex $r(T_2)$ is mapped to a lower point in $\Gamma'$ that the point it was mapped in $\Gamma$. The root $r(T_2)$ of subtree $T_2$ was mapped to point $p^L_{i_2-|upper(T_2)|}$ in $\Gamma$. Since $\Gamma$ does not respect the proper ordering $\lambda$, it holds that $T_2$ appears before $T_1$ in $\lambda$ and, thus, $|upper(T_2)| \leq |upper(T_1)|$. So, in $\Gamma'$ vertex $r(T_1)$ is mapped to a point that is at or below the one vertex $r(T_2)$ was mapped in $\Gamma$. We conclude that edge $(r(T_1), r)$ is upward in $\Gamma'$ and, thus, $\Gamma'$ is a restricted UPSE.

Be repeatedly identifying pairs of subtrees that cause a restricted UPSE drawing to not respect $\lambda$ and by transforming the drawing as described above, we can obtain a restricted UPSE drawing for tree $T$ on $S$ that respects the proper ordering $\lambda$ of the subtrees of $T$. □

## B  Outerplanar digraphs

In this section we extend our approach to the class of outerplanar digraphs. For better understanding we keep the new definitions on outerplanar graphs as close as possible to already existing definitions for trees.

Consider an *acyclic outerplanar digraph* $G$ (*outerplanar-DAG* for short), i.e., a directed acyclic graph whose underlying undirected structure is that of an outerplanar graph. A *cut vertex* is any vertex of $G$ that when removed increases the number of connected components. A maximal biconnected subgraph of $G$ is called a *block* of $G$. A vertex of $G$ which is either a source or a sink is referred to as a *switch*. Let $B$ be a block of $G$. Binucci *et al.* [3] proved the following lemma.

**Lemma 5 (Binucci *et al.* [3]).** *Let $G$ be a n-vertex DAG containing a k-vertex cycle-DAG $C$, for some $k \leq n$. Suppose that $C$ has at least two vertices $u$ and $v$ that are sources in $C$. Then there exists a convex point set $S$ of size $n$ such that $G$ has no upward straight-line embedding into $S$.*

This result easily extends to an arbitrary two-sided convex point set:

**Lemma 6.** *Let $G$ be a n-vertex outerplanar-DAG, let $B$ a block of $G$ and $S$ be a convex point set of size $n$. If $B$ contains either two sources or two sinks then $G$ does not admit an UPSE into $S$.* □

Since each block has an equal number of sources and sinks, thus, we can assume that every block of $G$ contains exactly one source and one sink. Next note that if there is an UPSE of an outerplanar-DAG $G$ into a convex point set $S$, then it is outerplane and thus in the following we consider only outerplane embeddings of $G$. If an outerplane embedding of a block $B$ is bounded by two paths one of which is a single edge, then $B$ is called a *one-sided block*, otherwise it is called a *two-sided block*. A vertex of $B$ which is not a switch of $B$ is called a *side-vertex* of $B$. The two paths between two switches of $B$ are called the *sides* of $B$. By $L(B)$ and $R(B)$ we denote the side-vertices of two different sides of a block $B$.

The edges of $G$ that do not belong to any biconnected component are supposed to be the *trivial blocks* of $G$. Thus the two end vertices of such edge are considered to be a source and a sink of the corresponding trivial block.

Let $v$ be a cut vertex of $G$. The blocks of $G$ that have $v$ as their switch are called *extremal blocks* of $v$, the remaining blocks incident to $v$ are called *side blocks* of $v$. By $b^-(v)$ (resp., $b^+(v)$) we denote the number of extremal blocks of $v$ that have $v$ as their sink (resp., source). By $b(v)$ we denote the total number of extremal blocks that contain $v$, i.e., $b(v) = b^-(v) + b^+(v)$.

Let $G$ be a *rooted outerplanar-DAG* if one of its vertices, denoted by $r(G)$ is designated as its *root*. We then say that $G$ *is rooted at* vertex $r(G)$.

Let $G_1^l, \ldots, G_{b^-(r)}^l, G_1^h, \ldots, G_{b^+(r)}^h$ be the rooted subgraphs of $G$ obtained by a cut at $r$ and having $r$ as their sink or source, respectively. They are called the *extremal subgraphs of $T$*. A subgraph of $G$, obtained by a cut at $r$ and having $r$ as its side-vertex is called a *side subgraph* of $G$. Note that the superscripts "$l$" and "$h$" indicate whether a particular subgraph of $G$ has $r$ as its sink or as its source, respectively.

The rooted subgraph of $G$ consisting of $G$'s root, $r$, together with $G_1^l, \ldots, G_{b^-(r)}^l$ is called the *lower subgraph of $G$*. The lower subgraph of $G$ is denoted by $lower(G)$ (Figure 5). Similarly, the rooted subgraph of $G$ consisting of $G$'s root, $r$, together with $G_1^h, \ldots, G_{b^+(r)}^h$ is called the *upper subgraph of $G$* and is also rooted at $r$. The upper subgraph of $G$ is denoted by $upper(G)$. Let a block of $G$ that contains its root $r$ and let $v$ be a vertex of this block, different from $r$. Let $r$ be a cut vertex and consider a cut at $v$. By $G(v)$ we denote the union of the connected components that do not contain $r$ (Figure 5).

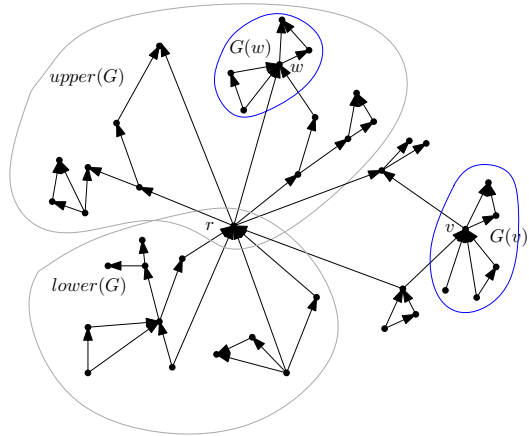

**Fig. 5.** An outerplanar-DAG $G$. A lower and an upper subgraphs of $G$ are surrounded by gray curves. The subgraphs $G(v)$ and $G(w)$ of graph $G$ are surrounded by blue curves.

Given two vertices $s$ and $t$ of $G$, a cut vertex $c$ of $G$ is called $(s,t)$-*separating* if the removal of $c$ leaves $s$ and $t$ in different connected components.

## B.1 Structure of outerplanar graphs that may admit an UPSE into a convex point set

In this section we elaborate on the structure of the outerplanar-DAGs that can have an UPSE into a convex point set.

**Lemma 7.** *Let $G$ be a $n$-vertex outerplanar-DAG, $S$ be a convex point set of size $n$, $s$ be a source of $G$ and $t$ be a sink of $G$. If there is an UPSE of $G$ into $S$ so that $s$ and $t$ are mapped to $b(S)$ and $t(S)$ respectively, then the following statements hold:*

(1) *For every path $P_{s,t}$ from $s$ to $t$ and for every two-sided block $B$ of $G$, $P_{s,t}$ contains either all vertices of $L(B)$ or all vertices of $R(B)$.*
(2) *For each cut-vertex $c$ there exists at most one block $B$ so that $c$ is a side vertex of $B$. Moreover, if $c$ is a separating $(s,t)$ vertex, then $c$ is a single side-vertex of $B$ at its side.*

*Proof.* (1) Assume for the sake of contradiction that there exists a path from $s$ to $t$, $P_{s,t}$, and a block $B$ so that $P_{s,t}$ does not contain neither $L(B)$ nor $R(B)$ completely. Let $L' = L(B) \setminus P_{s,t}$ and $R' = R(B) \setminus P_{s,t}$. Thus, none of the $L'$ and $R'$ is an empty set. Note that in any UPSE of $G$ into $S$ the path $P_{s,t}$ splits $S$ on one-sided point sets. While the vertices of $L'$ and $R'$ have to be mapped to different sides of $S$, we get a crossing with $P_{s,t}$ and clear contradiction.

(2) Let first $c$ be a cut-vertex. Assume for the sake of contradiction that there are two blocks $B_1$ and $B_2$ so that $c$ is a side-vertex for both $B_1$ and $B_2$. Since $S$ is convex, in any UPSE of $G$ into $S$, the vertices of $B_1$ and $B_2$ represent a convex point set. On the other hand, in any upward planar drawing of $B_1$ and $B_2$ so that all the vertices lie on the boundary of the drawing, the vertex $c$ has to be drawn inside the convex hall of the rest vertices. A clear contradiction.

Now assume that $c$ is not the only side-vertex of $B_1$ on its side, i.e., there is another side-vertex $c'$ of $B_1$, which belongs to the same side of $B_1$. Note that $c'$ is not the vertex of $P_{s,t}$, because otherwise $c$ is not an $(s,t)$-separating vertex. Thus $P_{s,t}$ does not contain nether $L(B_1)$ not $R(B_1)$. A clear contradiction by Statement 1 of the lemma. □

## B.2 A restricted UPSE problem for outerplanar-DAGs

Analogously to the restricted UPSE for trees, in this section, we study a restricted UPSE problem for outerplanar graphs. This problem is later on used by our main algorithm which decides whether there exists an UPSE of a given outerplanar-DAG into a given convex point set.

**Definition 2.** *In a restricted UPSE problem for outerplanar-DAGs we are given an outerplanar-DAG $G$ rooted at $r$, so that $G$ contains at most one two-sided block, and if so then $r$ is a single side-vertex on one of its sides (see Figure 5 for a possible entry graph $G$ of the problem). We are also given a convex point* *set $S$, and a designated point $p \in S$. We are asked to decide whether there exists an UPSE of $G$ into $S$ such that (i) the root $r$ of $G$ is mapped to point $p$ and, (ii) each subgraph of $G$ (extremal or side) is mapped to consecutive points on the same side (either $L$ or $R$) of $S$ (excluding the vertex $r$).*

Since each of the extremal subgraphs of $G$ has to be drawn on a one-sided point set, we first test if it is possible to draw each of them on a one-sided point set separately. This can be done by exploring the Theorem 3 of Binucci et al.[3] (see also Theorem 6 below). Let $F$ be an extremal subgraph of $G$. In order to test whether $F$ has an UPSE into an one-sided point set we adopt the definition of *auxiliary tree* from in [3].

Let $\mathcal{T}(F)$ be the BC-tree of $F$. A node $\mu$ of the auxiliary tree $\mathcal{T}'$ corresponds to a connected subtree $S$ of $\mathcal{T}(F)$ which is maximal with respect to the following property: A cut-vertex $c$ that belong in S and is shared by two blocks of $F$ is a switch vertex for both of them. An edge of $\mathcal{T}'(F)$, directed from $\mu$ to $\nu$ corresponds to a cut-vertex which is a side vertex for a block associated with $\mu$ and a switch vertex for a block associated with $\nu$.

The next theorem presents a necessary and sufficient condition for an outerplanar graph to have an UPSE into any one-sided point set.

**Theorem 6 (Binucci et al. [3]).** *A n-vertex connected DAG $G$ admits an upward straight-line embedding into every one-sided convex point set of size n if and only if the following conditions are satisfied:*

- *Condition 1: Every block of $G$ is regular.*
- *Condition 2: Every cut-vertex shared by two blocks is extremal for at least one of them.*
- *Condition 3: Every node of $\mathcal{T}'$ has at most one incoming edge[4].*

In the proof of Theorem 6 it is shown that if $r$ is a source (resp. sink) vertex of $F$ and belong to the vertex of $\mathcal{T}'(F)$ without incoming edges then $F$ has an UPSE into any one-sided point set so that $r$ is mapped to its lowest (resp. highest) point. We next show that this is also a necessary condition.

**Lemma 8.** *Let $F$ be an outerplanar-DAG without two-sided blocks and $r$ be a source (resp. sink) of $F$. Let $S_F$ be a one-sided point set consisting of $|F|$ points. If $r$ belongs to a vertex of $\mathcal{T}'(F)$ with a positive in-degree then $F$ does not admit an UPSE into $S_F$ so that $r$ is mapped to its lowest (resp. highest) point.*

*Proof.* Let $C_r$ be the set of blocks of $F$ that are represented in $\mathcal{T}'(F)$ by a single vertex and let $r \in C_r$. Since the vertex of $\mathcal{T}'(F)$, where $r$ belong to, has a positive indegree, there is a block $B$ in $F$, which is incident to $C_r$ by a vertex $v$, so that $v$ is a side-vertex of $B$. Let $v'$ be a vertex of $B$, connected to $v$ by edge $(v', v)$. In any UPSE of $F$ into $S_F$, the vertices of $C_r$ are mapped to the points of $S_F$ that are higher than the point where $v'$ is mapped to. Thus the lemma follows. □

---
[4] Without defining the tree $\mathcal{T}'$ this conditions sounds like: "Each cut-vertex of $G$ shared by several blocks is a non-switch vertex for at most one of them."

In case that an extremal subgraph $F$ of $G$, rooted at $r$ passes the test of Theorem [3] and, moreover $r$ belong to a vertex of $\mathcal{T}'(F)$ without incoming edges we infer that $F$ has an UPSE into any one-sided point set of $|F|$ points, so that $r$ is mapped to its extremal point and we call $F$ a *one-side embeddable* outerplanar-DAG.

In the next we assume that every extremal subgraph of $G$ is a one-side embeddable outerplanar-DAG.

Let again $G$, rooted at $r$, as described in the restricted UPSE problem. $G$ has at most one two-sided block. Moreover, if $B$ is such a block then $r$ is a unique side-vertex on one of its sides. Remind that $r$ is mapped to a pre-specified point $p$ of $r$. Consider an UPSE of $B$ into $S$, let $S_B$ be the consecutive points of $S$ used by $B \setminus r$. Let $S_B^l$ and $S_B^h$ be the point sets that are below and above $S_B \cup \{p\}$ respectively. Next we show how to test whether $lower(G)$ and $upper(G)$ admits an UPSE into $S_B^l \cup \{p\}$ and $S_B^h \cup \{p\}$.

W.l.o.g consider $upper(G)$ and $S_B^h \cup \{p\}$, the procedure for $lower(G)$ and $S_B^l \cup \{p\}$ is similar. We construct a tree $T_{upper(G)}$, called *upward skeleton* of $upper(G)$ and we prove that $upper(G)$ has a restricted UPSE into $S_B^h \cup \{p\}$, with $r$ mapped to $p$, if and only if $T_{upper(G)}$ has a restricted UPSE into $S_B^h \cup \{p\}$, with $r$ mapped to $p$. Let $G_1^h, \ldots, G_{b^+(r)}^h$ be the extremal subgraphs of $upper(G)$, see also Figure 6.a. We construct tree $T_{upper(G)}$ to consist of a root $r$ and subtrees $T_1^h, \ldots, T_{b^+(r)}^h$, rooted at vertices $r_1, \ldots, r_{b^+(r)}$, respectively, that are connected to $r$ by an outgoing from $r$ edges $(r, r_i)$, $1 \leq i \leq b^+(r)$, respectively. Consider $G_i^h$, let $v$ be the vertex incident to $r$, so that there is no directed path from $r$ to $v$ in $upper(G)$, except of the edge $(r, v)$. Recall that $G_i^h(v)$ is the union of extremal components attached to $v$ except the one which contains $r$ (see Figure 6.a, $G_i^h(v)$ is denoted by a blue curve). Tree $T_i^h$ consists of its root $r_i$ a directed path of length $|lower(G_i^h(v))| - 1$, having $r_i$ as its sink, and of a directed path of length $|G_i^h| - |lower(G_i^h(v))| - 1$, having $r_i$ as its source (see Figure 6.b).

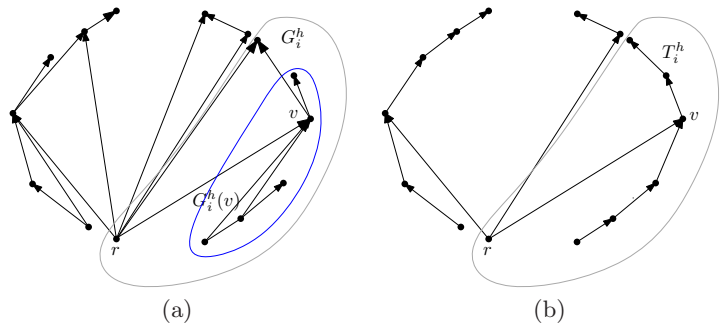

**Fig. 6.** (a) The graph $upper(G)$, its extremal subgraph $G_i^h$ and its subgraph $G_i^h(v)$. (b) The upward skeleton on $upper(G)$. The subgraph $lower(G_i^h(v))$ is substituted by a path of length $|lower(G_i^h(v))| - 1$.

**Lemma 9.** *The outerplanar-DAG upper$(G)$, rooted at $r$, admits a restricted UPSE into $S_B^h \cup \{p\}$ with $p$ mapped to $r$ iff each extremal subgraph of upper$(G)$, rooted at $r$ is one-side embeddable and upward skeleton $T_{upper(G)}$ of upper$(G)$ admits a restricted UPSE into $S_B^h \cup \{p\}$ with $p$ mapped to $r$.*

*Proof.* Assume that $upper(G)$, rooted at $r$, admits a restricted UPSE into $S_B^h \cup \{p\}$ with $p$ mapped to $r$. Let $\Gamma$ be such an embedding. We reconstruct this embedding into an UPSE of the $T_{upper(G)}$ into $S_B^h \cup \{p\}$. We consider $G_i^h$, $i = 1, \ldots, b^+(r)$. Since $\Gamma$ is a restricted UPSE, $G_i^h$ (except maybe vertex $r$) uses the consecutive points of $S_B^h \cup \{p\}$. Let $v$ be the vertex of $G_i^h$ incident to $r$, so that there is no other directed path from $r$ to $v$ in $G_i^h$ and let $G_i^h(v)$. In $T_i^h$ the $lower(G_i^h(v))$ is substituted by a directed path of length $|lower(G_i^h(v))| - 1$. We draw this path at the points where $lower(G_i^h(v))$ is drawn, using the consecutive points of $S_B^h \cup \{p\}$. The rest vertices of $T_i^h$ are drawn on the points where the remaining vertices of $G_i^h$ were placed. Now consider drawing of a single $G_i^h$, $i = 1, \ldots, b^+(r)$. $G_i^h$ is rooted at $r$ and $r$ is mapped to $p$. Let $S_i \cup \{p\}$ be the points of $S_B^h \cup \{p\}$ occupied by the $G_i^h$. Consider a virtual point $p_i$ that is below the points of $S_i$ so that $S_i \cup \{p_i\}$ creates a one-side convex point set. Place vertex $r$ to point $p_i$. Note that the new drawing of $G_i^h$ is upward and planar. Thus $G_i^h$ is one-side embeddable.

Now assume that $T_{upper(G)}$ admits a restricted UPSE into $S_B^h \cup \{p\}$ with $p$ mapped to $r$. Let $T_i^h$, $i = 1, \ldots, b^+(r)$ be the subtrees of $T_{upper(G)}$ rooted at $r_1, \ldots, r_{b^+(v)}$ respectively. Each of $T_i^h$, $i = 1, \ldots, b^+(r)$ is mapped to consecutive points of $S_B^h$, call this point set $S_i$. Now consider $G_i^h$, $i = 1, \ldots, b^+(r)$. Recall that $G_i^h$ is one-side embeddable. Consider $S_i \cup \{p_i\}$, where $p_i$ is a virtual point lower than all points of $S_i$. Map $G_i^h$ to $S_i \cup \{p_i\}$ so that $r$ is mapped to its lowest point, i.e. to $p_i$. Move $r$ to the point $p$. Note that in any UPSE of $G_i^h$ to $S_i \cup \{p_i\}$, vertex $v$ is mapped higher than all vertices of $lower(G_i^h(v))$. Thus, since $(r, v)$ is upward in the UPSE of $T_i^h$ it is also upward in the drawing of $G_i^h$ to $S_i \cup \{p\}$, where $r$ is mapped to $p$. □

**Theorem 7.** *Let $G$ be a n-vertex outerplanar-DAG rooted at vertex $r$, so that $G$ contains at most one two-sided block $B$, and if so then $r$ is a single side-vertex on one of its sides. Let $L$ and $R$ be left-sided and right-sided convex point sets, resp., such that $S = L \cup R$ is a convex point set of size $n$, and $p_r$ a point of $S$. The restricted UPSE problem with input $G$, $S$ and $p_r$ can be decided in $O(b(r)n)$ time. Moreover, if a restricted UPSE for $G$, $S$ and $p_r$ exists, it can also be constructed in $O(b(r)n)$ time.*

*Proof.* Let $B$ be a two-sided block of $G$ if there is one. W.l.o.g. assume that $L(B) = \{r\}$ and $p_r \in L$. Recall that in restricted UPSE an extremal subgraph of $G$ has to be drawn on consecutive points of $S$, thus if $u$, $v$ are vertices of $lower(G)$ and $upper(G)$ respectively, and both $u$ and $v$ are mapped to the same side of $S$, then $u$ has to be mapped lower than $v$. It is also clear that all points of $L$ below (resp. above) $p_r$ can be used only by $lower(G)$ (resp. $upper(G)$). Let $S_B^h$ be the subset of $|upper(G)|$ points of $S$ that is comprised by all the points of $L$ above $p_r$ (including) and consecutive highest points of $R$. Similarly, let $S_B^l$ be the subset of $|lower(G)|$ points of $S$ that is comprised by all the points of $L$ below $p_r$ (including) and consecutive lowest points of $R$. By the previous discussion in every restricted UPSE of $G$ into $S$, $upper(G)$ and $lower(G)$ are mapped to the points $S_B^h$ and $S_B^l$, respectively. Let now $S_B = S \setminus (S_B^l \cup S_B^l) \cup \{p_r\}$. The side subgraph $G_B$ of $G$ containing $B$ has to be mapped to the points of $S_B$. Next we prove that the existence of an UPSE of $G_B$ into $S_B$, so that $r$ is mapped to $p_r$, can be tested in linear time. Note that $S_B$ contains $|B| - 1$ consecutive points of $R$ and that all vertices of $R(B)$ are mapped to a one-sided point set. Thus, no vertex of $R(B)$ can be a side-vertex for any other block. I.e. for each $v \in B$, different from $r$, $B$ is the only block that has $v$ as its side-vertex. Note that in any UPSE of $G_B$ into $S_B$, the $upper(G_B(v))$ and $lower(G_B(v))$ have to be drawn above $v$ and below $v$ on the consecutive points next to $v$. Thus, for any $v \in R(B)$, it is enough to test whether each of its extremal subgraphs, rooted at $v$, is one-side embeddable. This can be done by their single traversal. The positions of the vertices of $B$ are then uniquely defined by the numbers of vertices attached to them, therefore the upwardness of edges of $B$ can be tested in linear time.

By Lemma 9, $lower(G)$ admits an UPSE into $S_B^h$ if and only if each extremal subgraph of $upper(G)$, rooted at $r$ is one-side embeddable and upward skeleton $T_{upper(G)}$ of $upper(G)$ admits a restricted UPSE into $S_B^h$ with $p$ mapped to $r$. One-side embeddability of extremal subgraphs of $upper(G)$ can be tested by their single traversal and therefore in linear time. Finally, the existence of a restricted UPSE of $T_{upper(G)}$ into $S_B^h$ can be tested in $O(b^+(r) \min(|L \cap S_B^h|, |R \cap S_B^h|))$ time, by using the dynamic programming procedure explained in Section 3 (see also Theorem 1).

Thus the overall complexity is $O(b(r) \min(|L|, |R|))$. □

Similarly to case of trees, denote by $\mathcal{L}(G, L, R)$ the set of points $p \in L \cup R$ such that there exists a restricted UPSE of $G$ on $L \cup R$ so that the root of $G$ is mapped to $p$. The next theorem follows easily from Theorem 7 by testing each point of $L \cup R$ as a candidate host for $r(G)$.

**Theorem 8.** *Let $G$ be an $n$-vertex outerplanar-DAG rooted at vertex $r$, so that $G$ contains at most one two-sided block $B$, and if so then $r$ is a single side-vertex on one of its sides. Let $L$ and $R$ be left-sided and right-sided convex point sets, resp., such that $L \cup R$ is a convex point set of size $n$. Then, set $\mathcal{L}(G, L, R)$ can be computed in $O(b(r)n \min(|L|, |R|))$ time.* □

### B.3 Testing algorithm for outerplanar-DAGs

Let $S$ be a convex point set of $n$ points. Let $G$ be a $n$-vertex outerplanar-DAG and $s, t$ a source and a sink of $G$. Assume that there exists a path $P_{s,t}$ from $s$ to $t$, that fulfills the requirements of Lemma 7, otherwise we can infer that there is no UPSE of $G$ into $S$, so that $s$ and $t$ are mapped to $b(S)$ and $t(S)$, respectively. Let $P^c_{s,t} = (s = w_1, \ldots, w_m = t)$ be the subpath of $P_{s,t}$ that contains only the cut-vertices which are also the $(s,t)$-separating vertices (see Figure 7.a). Every two consecutive vertices $w_i, w_{i+1}$ in $P^c_{s,t}$ belong to the same block of $G$, that can be also a trivial block, i.e. an edge. By $G_{w_i, w_{i+1}}$ we denote the graph that is induced by this block and all vertices connected to it by a path not passing through $w_i$ or $w_{i+1}$. The graphs $G_{w_1, w_2}, \ldots, G_{w_{m-1}, w_m}$ are called the *path-components* of $G$ defined by the path $P_{s,t}$.

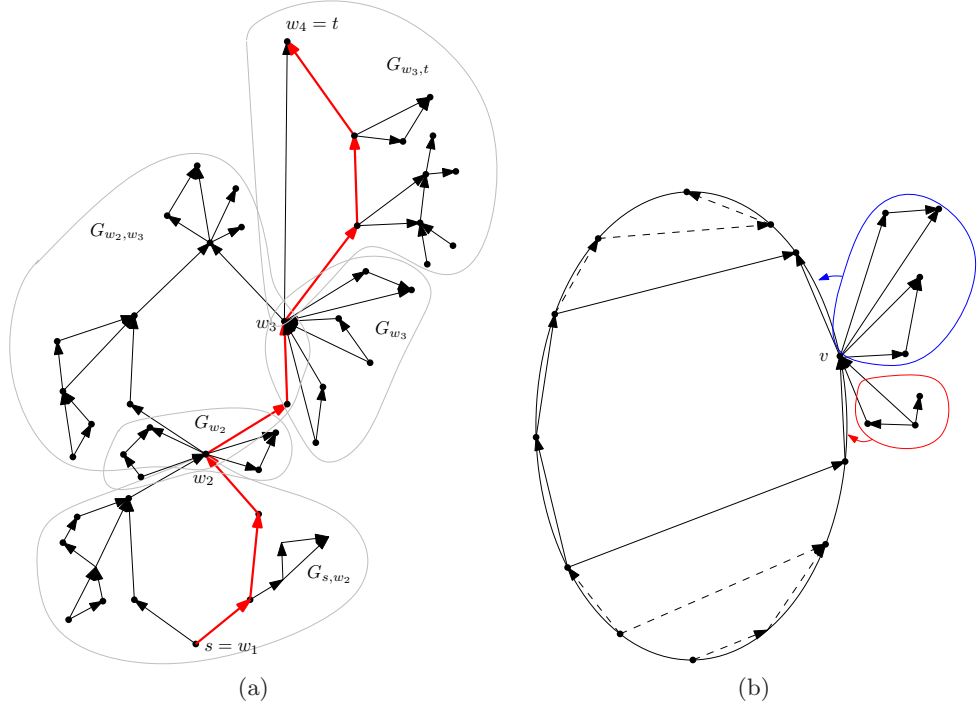

**Fig. 7.** (a) An path from $s$ to $t$ that fulfills the Lemma 7 is denoted by red. The subpath $P^c_{s,t} = \{s = w_1, w_2, w_3.w_4 = t\}$ are the cut vertices that are also $(s,t)$-separation vertices of $G$. The path-components $G_{w_i, w_{i+1}}$, $i = 1, \ldots, 3$ and the components $G_{w_i}$, $i = 1, \ldots, 4$ are denoted by grey curves. (b) An UPSE of $G_{w_{i-1}, w_i}$ and of path $P_{s,t}$ on some points of $S$. Path $P_{s,t}$ is denoted by dashed lines.

Consider the vertex $w_i$. Let a subgraph of $G$ that contains $w_i$ and is produced by deletion of edges of $G_{w_{i-1}, w_i}$ and $G_{w_i, w_{i+1}}$ that are incident to $w_i$. Denote this graph by $G_{w_i}$. Finally let $G_{s, w_i}$ be the subgraph of $G$ induced by the vertices of $G_{w_1}, \ldots, G_{w_i}$ and path-components $G_{s, w_2}, \ldots, G_{w_{i-1}, w_i}$. By definition, $G_{s, w_m} = G$ and $G_{w_1} = G_{s, w_1}$. By Lemma 1, each spanning tree of $G_{s, w_i}$ and therefore $G_{s, w_i}$ is drawn on consecutive points of $S$, call this point set $S_i$. Since $s$ is mapped to $b(S)$, we infer that $b(S) \in S_i$. Similarly, in any UPSE of $G$ into $S$, $G_{s, w_{i+1}}$ is also drawn on consecutive points of $S$ that contain $b(S)$, call this point set $S_{i+1}$. Hence, $G_{w_{i+1}}$ and path-component $G_{w_{i-1},w_i}$ are drawn on a set $S_{w_{i+1}} = S_{i+1} \setminus S_i$, that is, a subset of $S$ comprised by two consecutive point sets of $S$, one on its left and one on its right side. Next lemma elaborates on the possible drawings of a path-component $G_{w_{i-1},w_i}$, $i \leq m$.

**Lemma 10.** *Let $G$ be a n-vertex outerplanar-DAG and let $S$ be a convex point set of size $n$. Let $P_{s,t}$ be a path from a source $s$ to a sink $t$ of $G$. Let $G_{w_{i-1},w_i}$, $i \leq m$, be one of the path-components defined by the path $P_{s,t}$. Let $v$ be a vertex of the block of $G_{w_{i-1},w_i}$ that contains $w_{i-1}$ and $w_i$, but different from $w_{i-1}$ and $w_i$. In any UPSE of $G$ into $S$ such that $s$ and $t$ are mapped to $b(S)$ and $t(S)$ respectively, the vertices of $G_{w_{i-1},w_i}(v)$ are drawn on the same side of $S$ where $v$ is mapped and on the consecutive points of $S$ around $v$.*

*Proof.* Let $B$ be the block of $G_{w_{i-1},w_i}$ that contains both $w_{i-1}$ and $w_i$. Consider an UPSE of $B$ and of path $P_{s,t}$ on some points of $S$ (see Figure 7.b). Note that $P_{s,t}$ and $B$ split $S$ into one-sided point sets. Let $v$ be a vertex of $B$. Consider the subgraph $G_{w_{i-1},w_i}(v)$ of $G_{w_{i-1},w_i}$. By the previous observation the vertices of $G_{w_{i-1},w_i}(v)$ are drawn on a one-sided subset of $S$. If $G_{w_{i-1},w_i}(v)$ does not use the consecutive points around $v$, then the points which are left free are used by other vertices and since the graph is connected, crossings are introduced. Thus the lemma follows. □

From the previous lemma we infer that for every path-component $G_{w_{i-1},w_i}$ of $G$ it is sufficient to test only two of all its upward planar embeddings, the one of which is a mirrored image of the other.

Similarly to the case of trees, the dynamic programming Algorithm 2 maintains a list of points $\mathcal{P}(a,b,k)$, $0 \leq a \leq |L|$, $0 \leq b \leq |R|$, $1 \leq k \leq m$, such that:

$$p \in \mathcal{P}(a,b,k) \iff \begin{cases} G_{s,w_k} \text{ has an UPSE into point set } S_{1..a,1..b} \text{ with} \\ \text{vertex } w_k \text{ mapped to point } p. \end{cases}$$

The following theorem can be proved along the same lines as Theorem 3.

**Theorem 9.** *Let $G$ be a n-vertex outerplanar-DAG, $S$ be a convex point set of size $n$, $s$ be a source of $G$ and $t$ be a sink of $G$. Algorithm 2 decides in polynomial time whether $G$ has an UPSE on $S$ such that $s$ is mapped to $b(S)$ and $t$ is mapped to $t(S)$. Moreover, if such an UPSE exists, it can also be constructed in polynomial time.*

**Algorithm 2:** OUTERPLANAR-UPSE$(G, S, s, t)$

**input** : An outerplanar-DAG $G$, a point set $S$, a source $s$ and a sink $t$ of $G$ and a path $P_{s,t}$. Path $P^c_{s,t} = (s = w_1, \ldots, w_m = t)$ is used to progressively build graph $G$ from subgraphs $G_{w_{i-1},w_i}$, $2 \leq i \leq m$ and $G_{w_i}$, $1 \leq i \leq m$.

**output** : "YES" if $G$ has an UPSE into $S$ with $s$ mapped to $b(S)$ and $t$ mapped to $t(S)$, "NO" otherwise.

1. **For** $a = 0 \ldots |L|$
2.     **For** $b = 0 \ldots |R|$
3.         $\mathcal{P}(a, b, 1) = \mathcal{L}(G_{w_1}, L_{1..a}, R_{1..b})$
4.         **For** $k = 2 \ldots m$     //Consider outerplanar-DAG $G_{w_k}$
5.             $\mathcal{P}(a, b, k) = \emptyset$
6.             **For** $i = 0 \ldots |G_{w_k}|$     //We consider the case where $i$ vertices of $G_{w_k}$ are placed into the left side of $S$
7.                 **if** $(a - i \geq 0)$ **and** $(b - (|G_{w_k}| - i) \geq 0)$
8.                     Let $\mathcal{L} = \mathcal{L}(G_{w_k}, L_{a-i+1..a}, R_{b-(|T_{w_k}|-i)+1..b})$
9.                     //We consider all possible placements of $w_{k-1}$
10.                    //And first the case when the vertices $L(G_{w_{k-1}})$ are mapped to
11.                    //the left side of $S$
12.                 **For** each $q$ in
13.                     $\mathcal{P}(a - i - |L(G_{w_{k-1},w_k})|, b - (|G_{w_k}| - i) - |R(G_{w_{k-1}})|, k - 1)$
14.                     //We consider all the possible placements of vertex $w_k$
15.                     **For** each $p$ in $\mathcal{L}$
16.                       **if** (the drawing of $G_{w_{k-1},w_k}$ so that $L(G_{w_{k-1},w_k})$ are
17.                         mapped to $L_{a-i-|L(G_{w_{k-1},w_k})|+1..a-i}$ and
18.                         $R(G_{w_{k-1},w_k})$ are mapped to
19.                         $R_{b-(|G_{w_k}|-i)-|(R_{w_{k-1},w_k})|+1..b-(|G_{w_k}|-i)}$ is upward)
20.                       **then** add $p$ to $\mathcal{P}(a, b, k)$.
21.                 //Next we consider the case when the vertices $R(G_{w_{k-1}})$
22.                 //are mapped to the left side of $S$
23.                 **For** each $q$ in
24.                     $\mathcal{P}(a - i - |R(G_{w_{k-1},w_k})|, b - (|G_{w_k}| - i) - |L(G_{w_{k-1}})|, k - 1)$
25.                     //We consider all the possible placements of vertex $w_k$
26.                     **For** each $p$ in $\mathcal{L}$
27.                       **if** (the drawing of $G_{w_{k-1},w_k}$ so that $L(G_{w_{k-1},w_k})$ are
28.                         mapped to $R_{b-(|G_{w_k}|-i)-|(L_{w_{k-1},w_k})|+1..b-(|G_{w_k}|-i)}$
29.                         and $R(G_{w_{k-1},w_k})$ are mapped to
30.                         $L_{a-i-|R(G_{w_{k-1},w_k})|+1..a-i}$ is upward)
31.                       **then** add $p$ to $\mathcal{P}(a, b, k)$.
32. **if** $\mathcal{P}(|L|, |R|, m)$ is empty **then return**("NO");
33. **return**("YES");